\documentclass[letterpaper]{IEEEtran}
\IEEEoverridecommandlockouts
\usepackage{cite}
\usepackage{amsmath,amssymb,amsfonts}
\usepackage{graphicx}
\usepackage{textcomp}
\usepackage{xcolor}
\usepackage{capt-of,etoolbox}
\usepackage{amsmath}
\usepackage{amsfonts}
\usepackage{graphicx}
\usepackage{multirow}
\usepackage{multicol}
\usepackage{lipsum}
\usepackage{cite}
\usepackage{url}
\usepackage{amssymb}
\usepackage{balance}
\usepackage{multicol,tabularx,capt-of}
\usepackage{multirow}
\usepackage{siunitx}
\usepackage{gensymb}
\usepackage{algorithm}
\usepackage[noend]{algpseudocode}
\usepackage{textcomp}
\usepackage{varwidth}			
\usepackage{booktabs}
\usepackage[export]{adjustbox}
\usepackage{xcolor,colortbl}
\usepackage{dblfloatfix}
\usepackage{float}
\usepackage{placeins}
\usepackage{caption}
\usepackage{subcaption}
\usepackage{mwe}   
\usepackage{tikz}
\def\checkmark{\tikz\fill[scale=0.4](0,.35) -- (.25,0) -- (1,.7) -- (.25,.15) -- cycle;}

\definecolor{Gray}{gray}{0.85}
\definecolor{LightCyan}{rgb}{0.88,1,1}

\def\BibTeX{{\rm B\kern-.05em{\sc i\kern-.025em b}\kern-.08em
    T\kern-.1667em\lower.7ex\hbox{E}\kern-.125emX}}
\begin{document}

\title{Smart Village: An IoT Based Digital Transformation 
}



\author{
\begin{tabular}[t]{ccc@{\extracolsep{1em}}}
Amit Degada & Himanshu Thapliyal & Saraju P. Mohanty \\
University of Kentucky & University of Kentucky  & University of North Texas\\
Lexington, KY, USA & Lexington, KY, USA & Denton, TX, USA \\
amit.degada@uky.edu & hthapliyal@ieee.org & smohanty@ieee.org \\
 
\end{tabular}
}


\maketitle

\begin{abstract}

Almost 46\% of the world's population resides in a rural landscape. Smart villages, alongside smart cities, are in need of time for future economic growth, improved agriculture, better health, and education. The smart village is a concept that improves the traditional rural aspects with the help of digital transformation. The smart village is built up using heterogeneous digital technologies pillared around the Internet-of-Thing (IoT). There exist many opportunities in research to design a low-cost, secure, and efficient technical ecosystem. This article identifies the key application areas, where the IoT can be applied in the smart village. The article also presents a comparative study of communication technology options.  



\end{abstract}


\section{Why smart villages are needed?} 

Approximately 60 Million people and 97\% of the landmass in the US is classified as rural \cite{USPopulation}. The three most populated countries China, India, and Indonesia have a considerable amount of population are living in villages. Some countries, e.g., Nepal, Cambodia, Chad, Uganda, Malawi, Rwanda, St. Lucia, and Solomon Island, etc. have more than three fourth of the population lives in rural areas. It is well-understood that the nature of the problem in villages is beyond the quality of life and requires utmost attention \cite{sen2020smart}. Therefore, the general assembly of the United Nations (UN) has adopted seventeen Sustainable Development Goals (SDG) towards the holistic development of every aspect of an individual's life \cite{UNSDG}.

On the other hand, the technological advancement in communication and computing technologies have transformed health, education, food, energy access, and the way we interact with our surroundings on a large scale. In recent years, the research community has focused on the development of the sustainable settlements of the communities, e.g., smart cities. The development of the smart villages, parallel to the smart cities, are the need of the time to bring both at the same level to sustain equilibrium in holistic development. 

\begin{figure}[htbp]
\centering
\includegraphics[trim=0cm 0cm 0cm 0cm, scale=0.39]{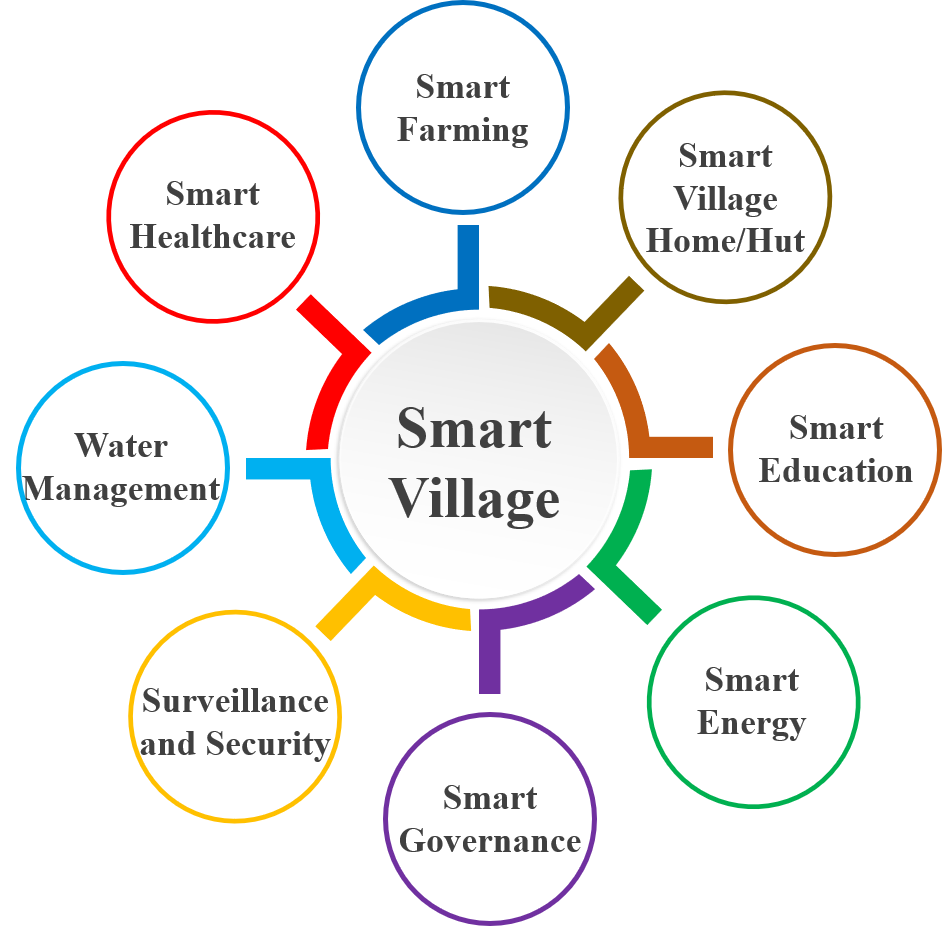}
\caption{Smart Village.}
\label{SmartVillageTheme}
\end{figure}

To envisage the smart village (Figure \ref{SmartVillageTheme}), the Internet-of-Thing (IoT) is a key technology player. The IoT provides interactive platforms for exchange of the information, and control the actions between the smart devices. Smart things are embedded with smart sensor-computing platforms and information exchanges across the communication protocols. The IoT is a proven solution with improved efficiency and reduction in time, as well as cost. This article aims to bridge the gap between the existing challenges in smart villages, and potential digital solutions through IoT technology. Wireless networking creates the skeleton for everything to communicate. The wireless technology in rural environments should be low power, large range, and have less bandwidth. The article presents a comparison between two competing technology, Long Range Wide Area Network (LoRaWAN) and Narrow Band-IoT (NB-IoT), enabling the designer to chose that can provide optimal performance in the smart village vertical.

\section{Smart Village design cycle}

\begin{figure*}[htbp]
\centering
\includegraphics[trim=0cm 0cm 0cm 0cm, scale=0.59]{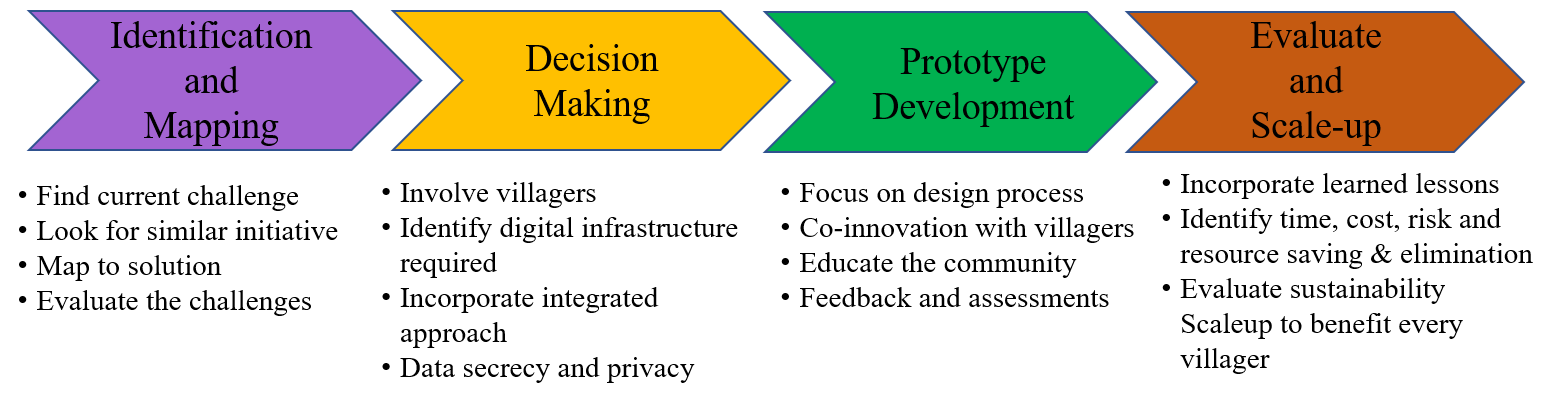}
\caption{Smart village design cycle.}
\label{svdesigncycle}
\end{figure*}

The setting up of the idea of smart communities is not recent. There are plenty of successful smart cities, and villages have already been established at many places across the globe. The smart village design uses smart technology to build viable services to alleviate the hardship of the village community, sustain the social ecosystem, and contribute towards economical growth. The smart village design problem is highly dispersed and often involves many non-hierarchical structures. Further, the resource in the villages are often shared, and a holistic approach is needed for sustainability. The four primary steps to design a smart village, as shown in Figure \ref{svdesigncycle} are identification and mapping, decision making, prototype development, and evaluation and scale-up. 

The first step is to identify the current problem, evaluate the challenges, and map it to a cost-effective smart digital solution for long-term inclusive growth. The second important step is to decide the relevant, appropriate and adaptable digital solutions to mobilize the resource and sustainable growth. The success of the smart village depends upon energy-efficient, low-cost power solutions, connected digital solutions, and uninterrupted connectivity. The key deciding factors of digital technology are coverage, data rate, power, and cost. The trust in user data privacy and secrecy are important for the success of the smart village. To ensure it, the design should specify how the data will be collected, used, stored, and shared. The third step is to deploy a scalable prototype. The prototype implementation helps to evaluate the deployed technology, study possible vulnerabilities, and explore another technological opportunity for better reach, revenue generation, and cost-effective solutions. The final step will replicate the prototype for the entire pool of villagers. This is often an iterative process to look for identifying the time, cost risk, and resource-saving or elimination. 

\section{IoT as backbone of the smart village}

\begin{figure}[htbp]
\centering
\includegraphics[trim=0cm 0cm 0cm 0cm, scale=0.35]{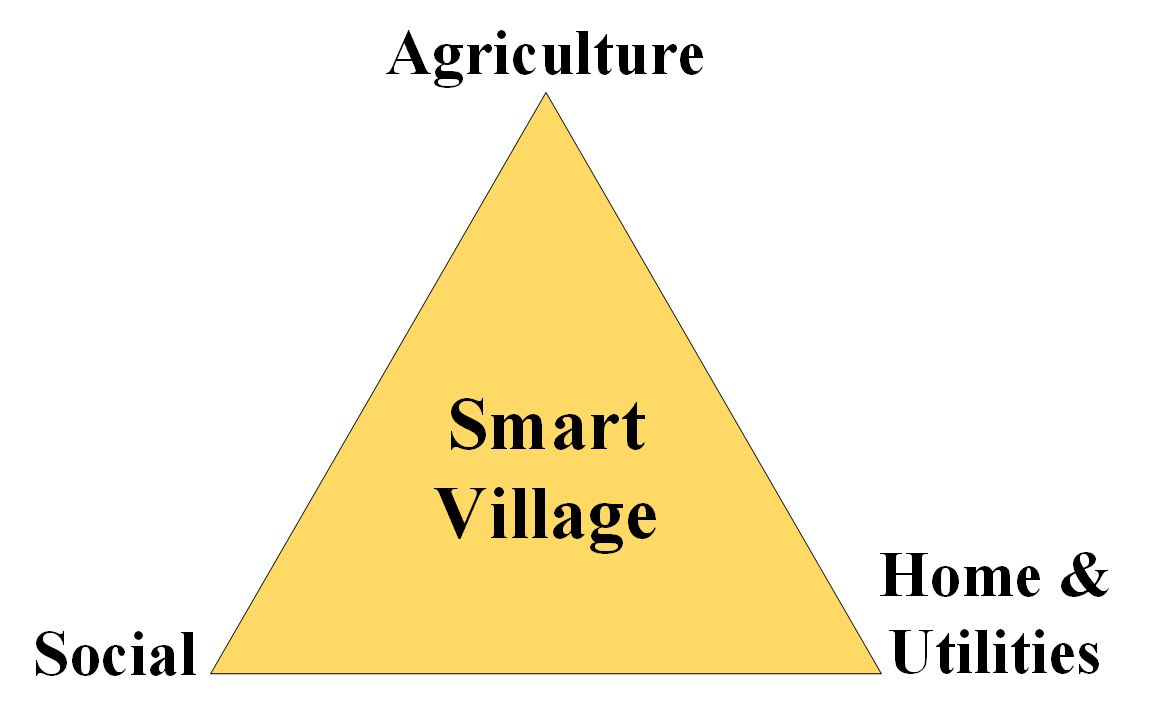}
\caption{Smart village key aspects.}
\label{Smartvillageaspects}
\end{figure}

The problem of designing the smart village can be complex and it should be addressed at multiple levels. The IoT application domain includes smart homes/huts, smart farming and dairy, smart cattle tracking, smart healthcare, smart water, and waste management, smart energy management, smart transportation, and smart governance, etc. We classify village-life aspects (Figure \ref{Smartvillageaspects}) into three categories: Agriculture,  Social, home, and utilities. Table \ref{SmartVillageClassification} summarizes the important issues being addressed in the literature so far.

\begin{table*}
\caption{The smart village goal} 
    \centering 
 \scalebox{1.00}{
    \begin{tabular}{ m{2cm}  m{5cm}   m{9.5cm} }
        \toprule
        Village-life aspects & Towards smart village  & Issue being addressed \\
        
        \toprule
        Agriculture & 
        \begin{itemize} 
            \item Smart farming
            \item Smart irrigation
            \item Smart live-stock tracking
            \item Smart dairy
            \item Smart agriculture waste management
        \end{itemize} & 

        \begin{itemize} 
            \item Climate monitoring and forecasting 
            \item Irrigation and fertilizer management system
            \item Crop health monitoring and predictive analysis
            \item Monitor the health and performance of individual animal and collective analysis as herd
            \item On-time forecasting and meeting the need of agriculture product 
            \item Reducing the agriculture waste and recycle management \cite{ Agriwastemanagement, LOEHR1974453}
        \end{itemize} \\
        \midrule
        
         Social & 
        \begin{itemize} 
            \item Education
            \item Surveillance and security
            \item Governance
            \item Infotainment
            
        \end{itemize} & 
       
        \begin{itemize} 
            \item Interactive learning experience, connected classrooms and libraries
            \item Understand  the learning style, attitude and adapt the methodology for individual student \cite{pruet2015exploring} 
            \item The real-time surveillance provide greater visibility, reach for prevention of crime, and critical asset loss
            \item Improve the governance efficiency, tracking the village wide services, safety of villagers, and enable sustainable growth \cite{anand2018governance}
        \end{itemize} \\
        \midrule
        
        \multirow{1}{*}{Home \& Utilities} & 
        \begin{itemize} 
            \item Smart village home/hut
            \item Smart healthcare 
            \item Smart Energy
            \item Water and waste management
        \end{itemize} & 
      
        \begin{itemize} 
            \item Gas, smoke-fire detection, camera surveillance and better energy management
            \item Catering the need of the elderly people, e.g health, stress and fall detection.  Monitor the chronic illness, e.g. high blood pressure, high cholesterol, diabetes, and rehabbing \cite{thapliyal2017smart}
            \item Security by design in energy harvesting consumer IoT devices \cite{ram2020energy}
            \item Efficient and cost-effective distribution of the clean water, check pH, oxygen, turbidity in water, hydrocarbons, metal and chemical contains in the soil \cite{ramesh2017water}
        \end{itemize} \\
        \midrule

    \end{tabular}
    \label{SmartVillageClassification} 
}
\end{table*}

Farming is perhaps the oldest industry, which requires the ability of the farmer to foresee the changes in surrounding nature, analyze the conditions and execute the required set of actions to increase the quality and quantity of the crops and livestock. IoT-based smart farming is seen as a prospect to revolutionize in many aspects. The IoT sensors collect weather information, soil moisture conditions, the health of crops and cattle. The ability to automate data analysis (using machine learning, big data) helps to better control the production cycle, growth, and meet a higher standard. The researchers in \cite{kamilaris2016agri}, proposed an online semantic framework that enables real-time data analytic, detects important events, automate decision making and works seamlessly among different sensors, process, applications, and operations. Similarly, IoT sensors can be used to track the location, monitor the health, and create a log of animal performance \cite{udutalapally2020scrop}.

In Europe and the US, villages have a large elder population. A significant portion of the village population is living below poverty. The people in the village are often deprived of good medical support and care at a low cost. IoT-based smart healthcare is a potential solution to provide evidence-based primary medical care. Chronic diseases (e.g. high blood pressure, high cholesterol, and diabetes) often cause heart and kidney disease, diabetes, blindness, and amputation. The authors in \cite{chui2017disease} reviewed several machine learning and emerging optimization algorithms for Alzheimer, dementia, tuberculosis, diabetes mellitus, and cardiovascular disease. They could be applicable for the smart village healthcare system.

The village usually adopts the traditional methods to collects the waste, thereby, lacks to create a healthy community. The waste produced in the village depends upon the agriculture season and varies widely depending upon the community. The IoT empowered solutions to help to identify the capacity of the trash cans, reduce overflows, determine the optimal route for garbage collection, and analysis of waste generation \cite{yusof2017smart, mahajan2017smart}. Smart energy management is meant to generate green electricity (e.g. Solar, windmill), monitor its usage, and control its wastage. The IoT sensors, controllers, solar panels, energy storage, and smart energy grids are key parameters for efficient and optimized usage. The IoT sensors placed at solar panels and windmills can sense the weather, and optimize electricity generation. Further, the smart energy system also manages the smart street light and electric vehicle charging. The extra energy generated from green energy sources (e.g. solar panels, windmills, etc.) can also sell extra energy to urban areas \cite{spanias2017solar,prinsloo2018synthesis }. The researchers in \cite{zhou2016big} have proposed a big data analytics model based on four key aspects, viz. management at power generation side, renewable energy, and its microgrid management, collaborative operations, and demand-side management. 

Other than limited employment, and revenue generation opportunity, the social aspects (e.g. education, safety, governance, etc.) is a common reason for village community migrations. The IoT-empowered education can track the attendance (using RFID, ZigBee), monitor the progress in quizzes (using cloud computing), smart libraries (with sensors), interactive learning experience, and resource-saving. Additionally, virtual reality-based smart education can be useful for vocational skill enhancement, healthcare training, and disaster management training, etc \cite{abdel2019internet, zhu2016research}. The smart education concept 
successfully made the school dropout to 0 with the introduction of smart education practices at model "smart village" Punsari, India \cite{singh2019achieving}.

\section{Building towards data driven ecosystem}

The digital world is driven by the collection of data efficiently, communication, and take intelligent decisions in a decentralized platform. The data-driven platform requires trusted information exchange between every "thing" in a secure, and efficient way. The foundation of the smart village depends upon choosing the right sensor, computing platform, energy supply mechanism, communication networking, and efficient information processing. The section provides an outline to build an IoT-based data-driven ecosystem for smart villages.

\subsection{IoT enabled digital transformation}
\begin{figure}[htbp]
\centering
\includegraphics[trim=0cm 0cm 0cm 0cm, scale=0.25]{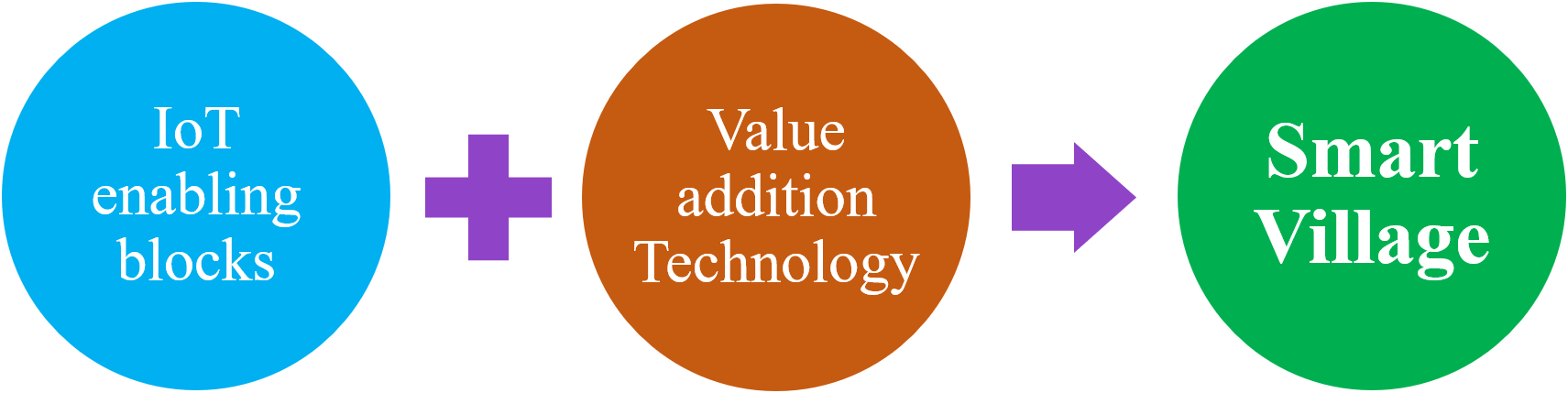}
\caption{Digital transformation towards smart village.}
\label{svdigitaltransformation}
\end{figure}

The IoT is an integration of sensors, actuators, and small computing units to automate a system via the internet. The digital transformation of a village is built up (as shown in Figure \ref{svdigitaltransformation}), primarily around IoT enabling blocks and value addition technology. The technologies that directly contribute are listed as follow:

\begin{itemize}

\item IoT Enabling blocks: The IoT is essentially machine-to-machine (M2M) communication protocols and interfaces. It includes the sensors, microcontrollers, energy harvesting mechanism, positioning technology (e.g. GPS), Radio Frequency Identification (RFID), Bluetooth low energy (BLE), Zig-Bee, Z-wave, Low Power Wide Area Networks (LPWANs), and Near-Field Communication (NFC).

\item Value addition technology: These technologies add the values, aid decision making, and enhance the security in overall service. The list includes cloud computing, Big data analytics, machine learning, artificial intelligence, virtual reality, and blockchain. Further, the usage of drones and robotics can also fall into this category.

\end{itemize}

\subsection{Wireless communication for wide-area network}

\begin{figure}[htbp]
\centering
\includegraphics[trim=0cm 0cm 0cm 0cm, scale=0.35]{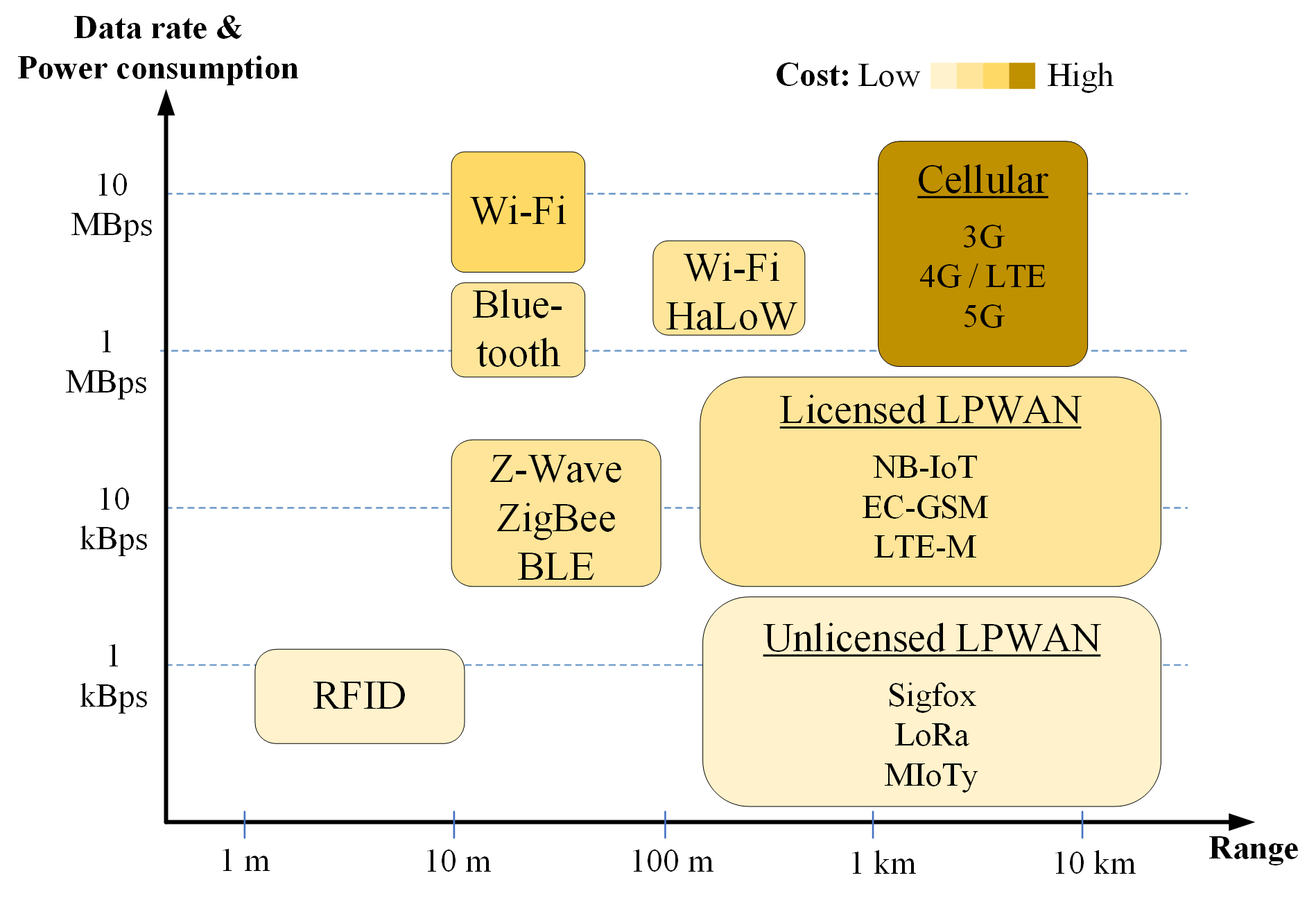}
\caption{Wireless communication technologies in IoT \cite{chanakIoT}\cite{WirelessIoT}.}
\label{WirelessIoT}
\end{figure}

The smart village deployment is often complex and suffers low connectivity. Thereby, no single communication technology can serve all the building blocks. The connectivity technology should support a wide area range, low power dissipation, support connectivity in all different climates, and can be set up quickly. The examples of wireless communication explored in literature are RFID, Bluetooth low power, Wi-Fi, NFC, LPWAN, and cellular technologies. Each smart village vertical has a unique set of requirements in terms of cost, range, power consumption, network management, and security. Table \ref{WirelessIoTAppn} maps the smart village verticals to preferred IoT communication technology. Further, Figure \ref{WirelessIoT} shows the comparison of different wireless technology in terms of distance, data rate, power consumption, and cost. 

\begin{table*}[htbp]
  \centering
  \caption{Mapping of smart village verticals to IoT communication technology.}
  \scalebox{1.00}{							
  \begin{tabular}{p{2.65cm} p{1.5 cm}p{1.5 cm}p{1.5 cm}p{1.5cm} p{1.5cm}p{1.5 cm}p{1.5 cm}}
  \toprule
  

   \textbf {\hfil Smart Village} & \hfil RFID & \hfil ZigBee & \hfil Z-Wave &\hfil  Bluetooth & \hfil  Wi-Fi & \hfil LPWAN & \hfil Cellular \\ 
   
   \textbf {\hfil verticals} &  &  &  & \hfil BLE  & \hfil  Wi-Fi Halow&  &  \\  \toprule

Climate monitoring in smart Farming & \hfil -- & \hfil -- & \hfil -- & \hfil -- &  \hfil -- & \hfil \checkmark & \hfil --  \\ \midrule
  
Irrigation &\hfil -- &\hfil -- &\hfil -- & \hfil -- & \hfil  -- & \hfil \checkmark & \hfil --  \\ \midrule
  
Livestock monitoring & \hfil \checkmark & \hfil -- & \hfil -- & \hfil -- & \hfil  -- & \hfil \checkmark & \hfil --  \\ \midrule

Dairy & \hfil \checkmark & \hfil \checkmark & \hfil -- & \hfil -- & \hfil  -- & \hfil \checkmark & \hfil --  \\ \midrule

Health-care & \hfil -- & \hfil -- & \hfil -- & \hfil \checkmark & \hfil  \checkmark & \hfil -- & \hfil \checkmark  \\ \midrule

Energy harvesting and management & \hfil -- & \hfil -- & \hfil -- & \hfil -- & \hfil  -- & \hfil \checkmark & \hfil --  \\ \midrule

Smart lighting & \hfil -- & \hfil -- & \hfil -- & \hfil -- & \hfil  -- & \hfil \checkmark & \hfil --  \\ \midrule 

Smart village home/hut & \hfil -- & \hfil \checkmark & \hfil \checkmark & \hfil \checkmark & \hfil  \checkmark & \hfil -- & \hfil \checkmark  \\ \midrule 

Surveillance  & \hfil -- & \hfil \checkmark & \hfil -- & \hfil -- & \hfil  \checkmark & \hfil -- & \hfil --  \\ \midrule 

Asset monitoring and tracking  & \hfil \checkmark & \hfil -- & \hfil -- & \hfil -- & \hfil  -- & \hfil \checkmark & \hfil \checkmark  \\ \midrule

Water and waste management  & \hfil \checkmark & \hfil -- & \hfil -- & \hfil \checkmark & \hfil  \checkmark & \hfil \checkmark & \hfil --  \\

  \bottomrule
  \end{tabular} 
   
  \label{WirelessIoTAppn}
  }						
\end{table*}

The majority of the smart village verticals do not require high speed and large bandwidth. Low Power Wide Area Network (LPWAN) technology emerged as the preferred wireless technology for M2M communication, offering cost-effective and energy-efficient solutions. The LPWAN uses an intermittent data packet size of  10 to 1000 bytes at a data rate ranging from 3 kbps to 375 kbps. Additionally, the LPWAN supports a large number of devices compared to other IoT communication technology. The first important choice that a designer needs to make is to use either unlicensed or licensed LPWAN. The unlicensed LPWAN uses the free license spectrum which could bring the cost down. However, there is the possibility that other organizations could use the same spectrum. It could result in interference from the different networks operating on the same spectrum. On the other hand, the licensed LPWAN supports a cellular communication network (e.g. 2G, 3G, 4G, and 5G). The licensed LPWAN facilitates instant seamless connectivity, the 'roaming' among networks, and operates on an exclusive spectrum allocated to a particular mobile operator. Further, it results in better secure, and reliable connections. 

\begin{figure}[h]
\centering
\includegraphics[trim=0cm 0cm 0cm 0cm, scale=0.45]{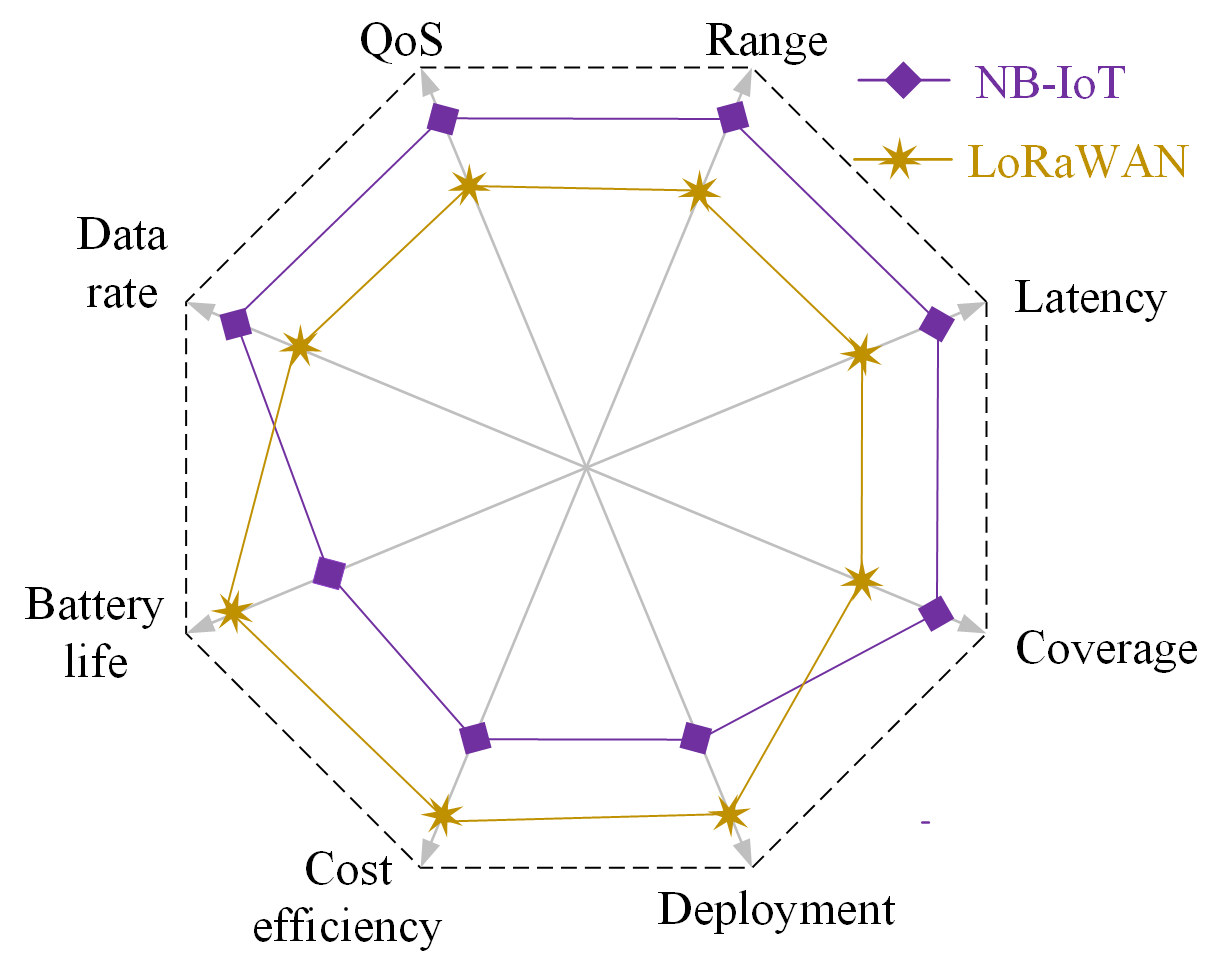}
\caption{LoRaWAN and NB-IoT Comparison \cite{sinha2017survey}.}
\label{LoRaNBIoTGraphCmp}
\end{figure}

The licensed LPWAN are based on cellular transmission, e.g. Narrow Band-IoT (NB-IoT), Long Term Evaluation-Machine type communication (LTE-M), Extended Coverage-GSM (EC-GSM). The unlicensed LPWAN uses spread-spectrum technology in Long Range WAN (LoRaWAN), telegram splitting in MIoTy, and ultra-narrowband in SigFox schemes. It is often to fall into the trap to believe that LoRa and LoRaWAN is interchangeable. The LoRa is a modulation technique based on Chirp Spread Spectrum (CSS), while LoRAWAN is a WAN protocol defined up to the application layer, and permitting the connection of several devices. Among licensed and unlicensed LPWAN: NB-IoT, and LoRaWAN are more suitable in smart village applications due to their bandwidth, data-rate, battery requirement, and mobility. Figure \ref{LoRaNBIoTGraphCmp} shows the comparison of LoRaWAN and NB-IoT. The selection of the technology depends upon IoT factors, i.e. range, cost, coverage, battery life, latency, and Quality of Service (QoS). LoRaWAN and NB-IoT have the respective advantage compared to different IoT factors. LoRaWAN is better in terms of coverage, deployment, cost efficiency, and battery life. However, NB-IoT is advantageous for range, data-rate, latency, and QoS. 

\begin{table}[htbp]
  \centering
  \caption{Comparison of LoRaWAN and NB-IoT \cite{sinha2017survey}.}
  \scalebox{1.00}{							
  \begin{tabular}{p{2.15cm} p{2.0 cm}p{2.0 cm}}
  \toprule
  

   \textbf {\hfil Technical} & \hfil LoRaWAN & \hfil NB-IoT\\ 
   
   \textbf {\hfil Parameter} &  &   \\  \toprule

\hfil Bandwidth & \hfil 125 kHZ & \hfil 180 kHZ  \\ \midrule
  
\hfil Coverage &\hfil 157 dB  &\hfil 164 dB  \\ \midrule
  
\hfil Security & \hfil AES & \hfil 3GPP \\ 
 & \hfil 128 bit  & \hfil (128 - 256 bits) \\ \midrule

\hfil Battery life &\hfil 15+ years  &\hfil 10+ years  \\ \midrule

\hfil Peak Current & \hfil 32 mA  & \hfil 12-130 mA  \\ \midrule

\hfil Sleep Current & \hfil 1 $\mu$A & \hfil 5 $\mu$A   \\ \midrule

\hfil Latency & \hfil latency & \hfil $<$ 10 S \\ 
 & \hfil insensitive & \hfil  \\ \midrule

\hfil Uplink MCL & \hfil 165 dB & \hfil 145-169 dB \\ \midrule 

\hfil Downlink MCL & \hfil 165 dB & \hfil 151 dB   \\ \midrule 

\hfil Range  & \hfil $<$ 15 KM & \hfil $<$ 35 KM  \\ \midrule 

\hfil Spectrum   & \hfil Free & \hfil $>$ 500 USD  \\  
 \hfil cost & \hfil  & \hfil  million/MHz \\  \midrule

\hfil Network and   & \hfil 100-1000 USD & \hfil 15K USD  \\ 
\hfil deployment cost  & \hfil per gateway & \hfil per base station  \\ 

  \bottomrule
  \end{tabular} 
   
  \label{LoRaNBIoTTechCmp}
  }						
\end{table}

There are several factors needed to be considered to select the best communication technology. Table \ref{LoRaNBIoTTechCmp} lists the comparison of NB-IoT and LoRa in terms of bandwidth, coverage, battery life, security, scalability, spectrum cost, and deployment cost, etc. However, there is no clear winner, and each technology has the potential to meet a specific smart village vertical. The superior choice depends upon the life cycles, coverage area, networking options, ability to access the device.

\section{Conclusion}

The smart village concept addresses many life verticals of rural life, involves multiple stakeholders, and should be self-sustaining. Agriculture, education, health, energy, water, and waste management, governance are the critical village life verticals to achieve sustainable development. The IoT is a digital transformation platform that can serve as the backbone to develop the digital data-driven ecosystem. The digital transformation requires IoT communication technology that can serve a large area, work on limited power, less setup time, and operates across different climates. The LoRaWAN and NB-IoT are seen as closely competitive options, and there is no clear winner. The comparison between LoRaWAN and NB-IoT presented in this article can provide help to smart village designers select the optimal communication technology.

\bibliographystyle{IEEEtran}
\bibliography{References}

\end{document}